\documentclass[a4paper, amssymb, amsmath, reprint, aps, nofootinbib, twoside, pra]{revtex4-2}
\usepackage[utf8]{inputenc}
\usepackage{graphicx}
\usepackage{physics}
\usepackage{hyperref}
\usepackage{mathtools}
\usepackage{xcolor}
\newcommand{\boldalpha}{\boldsymbol{\alpha}}

\begin{document}


\title{Nuclear Deformation Effects in the Spectra of Highly Charged Ions}

\author{Zewen Sun, Igor A. Valuev, Natalia S. Oreshkina}
\affiliation{Max-Planck Institute for Nuclear Physics, Saupfercheckweg 1, 69117 Heidelberg, Germany}


\begin{abstract}

Nuclear deformation effects are theoretically investigated in terms of deformation corrections of the electronic binding and transition energies, $g$ factor, and hyperfine splitting constant. 
By solving the Dirac equation twice, with the nuclear potential calculated from Fermi and deformed Fermi nuclear density distributions, we separate the deformation effect in binding energies and wavefunctions. 
The parameters for both models are determined from experimental data. 
The considered corrections are of interest for spectral analysis and are numerically calculated for the widest possible range of nuclei, consisting over 1100 different samples. 
The subtleties between different sources of measured data and the corresponding results are discussed. 
In addition, the importance of deformation effects for the search of new physics with singly-charged ions is examined.

\end{abstract}

\maketitle


\section{Introduction}
\label{sec:intro}
The simplicity of highly charged ions provides feasibility of probing nuclear properties with high precision. 
This ideal platform gives a wide scope for fundamental theoretical \cite{PRA.77.032501, PRL.108.063005, PRA.99.012505} and experimental research \cite{PRL.91.183001, PRL.94.223001, PRA.87.032714, PRA.91.042705}, making it a well-tested and understood system. 
Because of the smaller number of electrons and their closer distance to the nucleus compared to neutral atoms, highly charged ions are also more sensitive to relativistic and nuclear effects.
With the rapidly increasing precision of measurements, e.g., transition energies \cite{Heavner_2014} and $g$ factor \cite{PRL.97.030801, PRL.85.5308, PRL.92.093002, PRL.107.023002, PRL.122.253001, PRL.123.173001}, it is important to have a clear understanding when nuclear shape effects come into play for different observables and for different nuclei. 
The nuclear deformation (ND) effects are caused by nuclear shape distortion against a spherically symmetric distribution, which is one of the largest subleading nuclear structure effects \cite{DEBIERRE2020135527, PRA.106.062801}. 
Most of the nuclei are actually deformed, but the atomic theory predictions are almost exclusively based on spherically symmetric charge distributions.

From an experimental point of view, ND effects can become relevant either when the required precision is high, e.g., atomic clocks, or when the nuclear effects are sizable. 
The extraction of nuclear parameters, e.g., a relative change in root-mean-square (RMS) radii from King's linearity, has been commonly performed with significant simplifications of ND effects \cite{PhysRev.188.1916, FRICKE1995177, ANGELI201369}. 
Therefore, it is important to understand when and to which extent it is justified.
More importantly, detailed investigations of ND effects can illuminate the frontier between physics within and beyond the Standard Model (SM). 
There are proposals that point out the possibility of searching for a new boson  by observations of nonlinearity of the King plot \cite{PRD.96.093001, PRL.120.091801}.  
However, different opinions suggest that the nonlinearity can be explained within the SM, by considering ND effects \cite{PRA.103.L030801} or a quadratic field shift \cite{PRL.125.123002}. 
For both of the two opinions, the validity check requires highly accurate calculations \cite{PRX.12.021033}. 
The study of ND effects provides potential falsifiable evidence for the existence of physics beyond the SM, as well as insights into the study of isotope shift.

In this paper, we characterize the nuclear shape model by the RMS charge radius $R_\mathrm{RMS}$ and the intrinsic quadrupole moment $Q_0$.
We analyze the importance of the sign of $Q_0$, which defines the types of deformed nuclear shapes, but quite often is not indicated explicitly. 
In such cases, we follow the general assumption of $Q_0 \ge 0$, as well as emphasize the measured negative signs, when available.

We probe ND effects for the widest possible range of nuclei by theoretical calculations on a few observables: the binding and transition energies, $g$ factor, and hyperfine splitting constant. When possible, we also analyze and discuss the similarities and discrepancies between our results and previous studies on ND correction to $g$ factor~\cite{PRL.108.063005, PRA.99.012505}.
Finally, we discuss ND contribution in the spectra of singly-charged $\mathrm{Yb}^+$ ions as a source of King's plot nonlinearities. 

Relativistic units ($ c=m_e=\hbar=\epsilon_0=1 $) are used throughout the paper, unless explicitly given.


\section{Hamiltonian with deformed nuclei}
\label{sec:Hamiltonian}

The relativistic Hamiltonian obtained from Dirac equation is 
\begin{equation}
    H = \boldalpha \cdot \mathbf{p} + \beta + V(r), 
\end{equation}
where $\boldalpha$ and $\beta$ are the Dirac matrices, and $\mathbf{p}$ is the momentum operator of the electron, and the potential is defined as~\cite{PRA.46.3735}
\begin{equation}
\label{V(r)}
    V(r) = 
     -\frac{Z \alpha}{r} \left[ \int_0^{r} {r'}^2 \rho \left( r' \right) \, dr'
    + r \int_r^\infty r' \rho \left( r' \right) \, dr' \right], 
\end{equation}
where $Z$ is the nuclear charge number, $\alpha$ is the fine-structure constant, and $\rho(r)$ is a spherical charge density.

In the general case of a non-spherical nucleus, the nuclear charge is described by the deformed Fermi distribution
\begin{equation}
\label{dFermi}
    \rho_{c\beta_2} (r, \theta) = \frac{ \rho_0 }{1+ \exp( \frac{r-c [ 1+\beta_2 Y_2^0(\theta) ] }{a} )}, 
\end{equation}
where $c$, $\beta_2$, and $a$ are the parameters for the model, $ Y_2^0(\theta) $ is the spherical harmonic, and $\rho_0$ is the corresponding normalization factor. 
The spherically symmetric $\rho(r)$ in Eq.~\eqref{V(r)} can be obtained by averaging $\rho_{c\beta_2} (r, \theta)$ on the $\theta$ coordinate.

The parameters $c$ and $\beta_2$ are determined such that the nuclear charge radius $R_\mathrm{RMS}$ and the intrinsic nuclear quadrupole moment $Q_0$ calculated from $\rho_{c\beta_2} (r, \theta)$ match the tabulated nuclear data: 
\begin{align}
\label{RMS}
    R_\mathrm{RMS}^2 &= \int \rho_{c\beta_2} (r, \theta) \, r^2 \ d^3r, \\
\label{Q_0}
    Q_0
    &= Ze \int \left(3 \cos^2 \theta -1 \right) \rho_{c\beta_2} (r, \theta) \, r^2 \, d^3r. 
\end{align}
The commonly adopted value for parameter $a$ is 0.5234~fm \cite{PRA.46.3735}. 
It should be noted that for light nuclei, this value is comparable with their $R_\mathrm{RMS}$, which is clearly too large to deliver a reasonable description.
Therefore, we artificially decrease the parameter as $ a = 0.15 \, R_\mathrm{RMS}$ for the cases when $ R_\mathrm{RMS} < 1.5 \ \mathrm{fm} $. 

\begin{figure*}[t]
    \centering
    \includegraphics[width=0.9\textwidth]{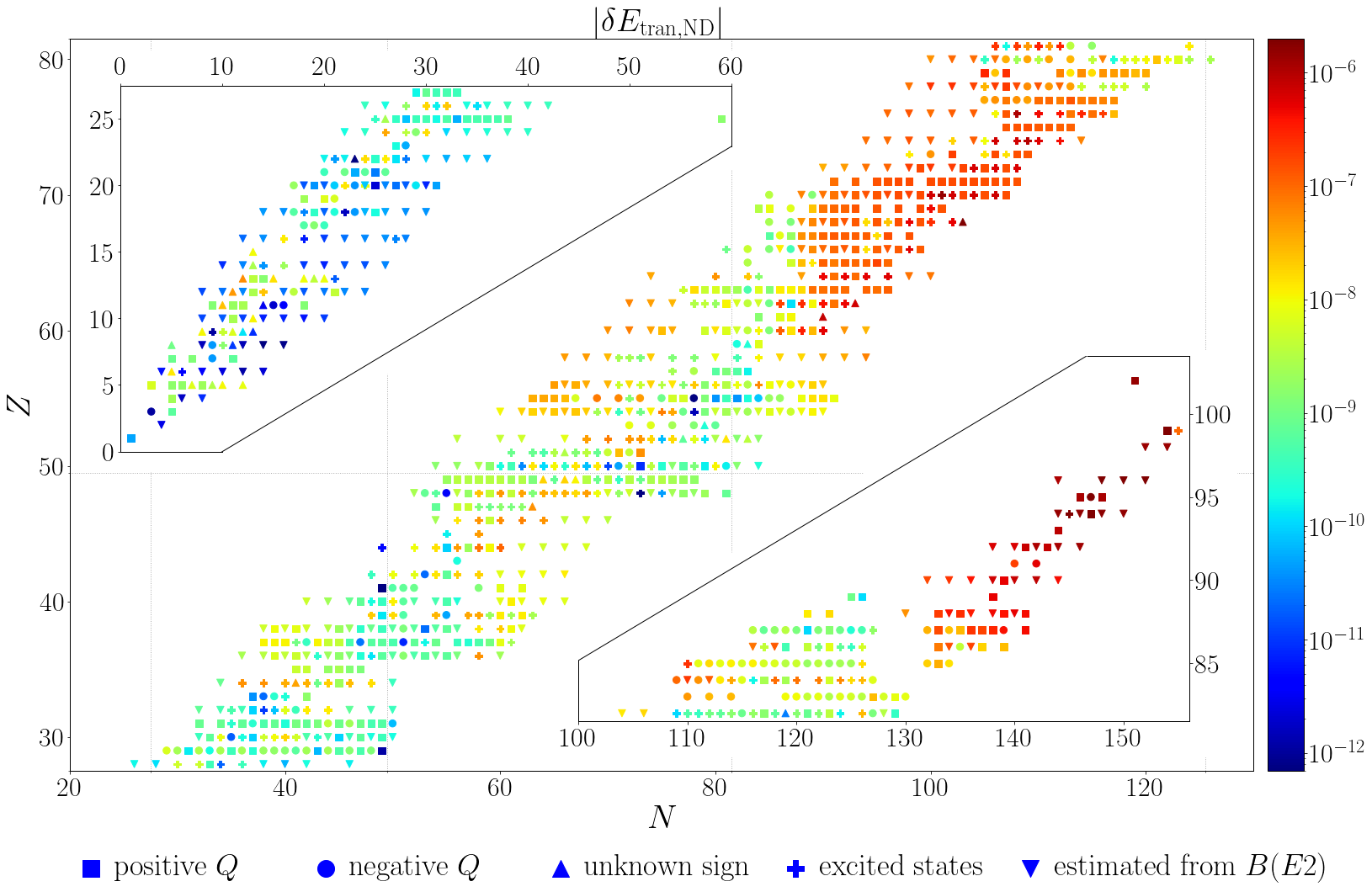} 
    \caption
    {\label{fig:1}Nuclear chart for ND  correction $\delta E_{\mathrm{tran,ND}}$ in $1s_{1/2} \rightarrow 2p_{1/2}$ transition energy. 
    The gray grid lines refer to the magic numbers on the periodic table (20, 28, 50, 82, and 126). The marker styles indicate the properties of $Q$ values used to calculate the ND correction, i.e., square: $Q$ is positive; circle: $Q$ is negative; up triangle: the sign of $Q$ is unknown but we take it as positive; cross: $Q$ obtained from a nuclear excited state instead of nuclear ground state; down triangle: $Q_0$ value is estimated from $B(E2)\uparrow$ transition rate (more explanation in Sec.~\ref{sec:discuss}). }
\end{figure*}

\begin{figure*}[t]
    \centering
    \includegraphics[width=0.9\textwidth]{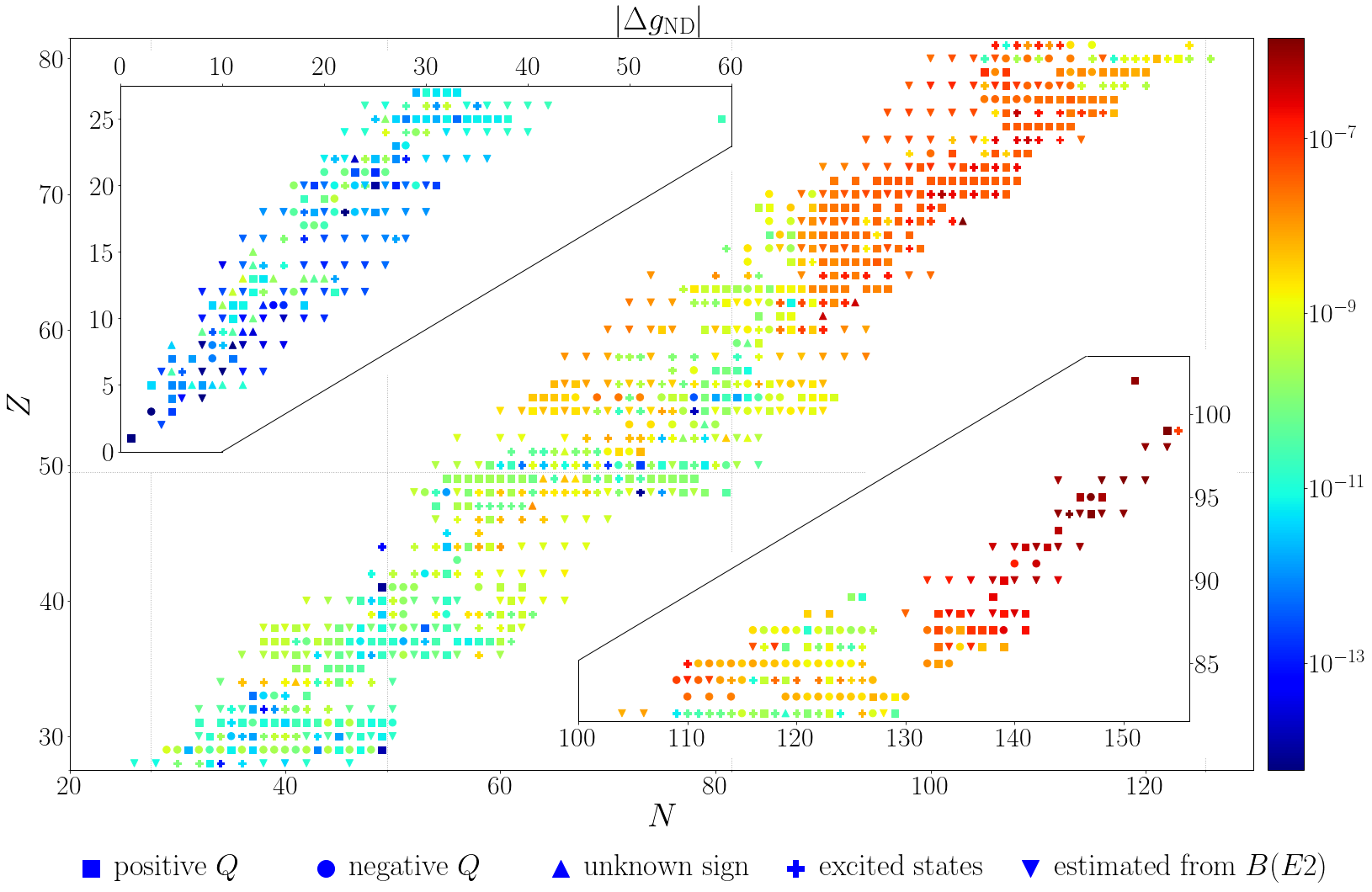} 
    \caption
    {\label{fig:2}Nuclear chart for ND $g$ factor correction $\Delta g_{\mathrm{ND}}$ in $1s$ state. See Fig.~\ref{fig:1} for the legend. }
\end{figure*}

\section{nuclear deformation corrections}
\label{sec:nuclear deformation corrections} 
In this paper, we investigate the ND effects in terms of transition energy $\delta E_{\mathrm{tran,ND}}$, and $g$ factor $\Delta g_{\mathrm{ND}}$. 
By numerically solving the Dirac equation
\begin{equation}
\label{H}
    H \Psi_{n \kappa m} (\mathbf{r}) = E_{n \kappa} \Psi_{n \kappa m} (\mathbf{r}), 
\end{equation}
we obtain the energies $E_{n \kappa}$ and the wavefunctions $\Psi_{n \kappa m}$, with
\begin{equation}
    \Psi_{n \kappa m} (\mathbf{r}) = \frac{1}{r} 
    \begin{pmatrix}
    G_{n\kappa} (r) \Omega_{\kappa m} (\theta, \varphi)\\
    iF_{n\kappa}(r) \Omega_{-\kappa m} (\theta, \varphi)
    \end{pmatrix}, 
\end{equation}
where $n$ is the principal quantum number, $\kappa$ is the relativistic angular momentum number, and $m$ is the total magnetic number. 

The correction $\delta E_{\mathrm{tran,ND}}$ is the relative difference between the transition energies of deformed $E_{\mathrm{tran}}^{(c\beta_2)}$ and non-deformed $E_{\mathrm{tran}}^{(c 0)}$ Fermi nuclear models
\begin{equation}
\label{dE}
    \delta E_{\mathrm{tran,ND}} = \frac{ E_{\mathrm{tran}}^{(c\beta_2)} - E_{\mathrm{tran}}^{(c0)} }{ E_{\mathrm{tran}}^{(c0)} }. 
\end{equation}
In our paper, we examine the transition energy between $1s_{1/2} \rightarrow 2p_{1/2}$ states.

The $g$ factor is calculated theoretically from the radial wavefunctions:
\begin{equation}
\label{g_fac}
    g = \frac{2\kappa}{j(j+1)} \int_0^\infty G_{n\kappa} (r) F_{n\kappa} (r) r \, dr, 
\end{equation}
where $j = \abs{\kappa} - 1/2$ is the total angular momentum number. 
The $\Delta g_{\mathrm{ND}}$ is ND correction to the ground-state electron $g$ factor:
\begin{equation}
\label{dg}
    \Delta g_{\mathrm{ND}} = g_{1s}^{(c\beta_2)} - g_{1s}^{(c0)}. 
\end{equation}

Additionally, ND corrections to $1s$ state eigenenergy $\delta E_{\mathrm{ND}}$ and hyperfine splitting constant $\delta \mathcal{A}_{\mathrm{ND}}$ are defined and discussed in the supplemental material \cite{supplement}.

Our calculations mainly consist of two parts. 
The first part is the calculation of parameters $c$ and $\beta_2$ based on the experimentally measured ground-state nuclear RMS radius and quadrupole moment. 
The radii $R_{\mathrm{RMS}}$ are obtained from Ref.~\cite{ANGELI201369}. 
Additionally, when the data are not available, we use the following empirical equation
\begin{equation}
    R_{\mathrm{RMS}} = \sqrt{\frac{3}{5}} R_0 = \sqrt{\frac{3}{5}} \left( 1.2 A^{1/3} \right) \ \mathrm{fm}, 
\end{equation}
where $R_0$ would correspond to the radius of a homogeneously charged sphere, and $A$ is the nuclear mass number. 
The intrinsic quadrupole moment $Q_0$ values are retrieved from Refs.~\cite{STONE2016, STONE2021, PRITYCHENKO2016}. 
Refs.~\cite{STONE2016, STONE2021} report spectroscopic nuclear quadrupole moment $Q$, which is related to $Q_0$ via
\begin{equation}
    Q_0 = \frac{(I+1)(2I+3)}{I(2I-1)} Q, 
\end{equation}
where $I$ is the nuclear spin quantum number. 
Ref.~\cite{PRITYCHENKO2016} provides adopted reduced quadrupole transition rates $B(E2)\uparrow$ for spinless nuclei from the nuclear ground state to the first excited $2^+$ state,  and $Q_0$ is obtained from
\begin{equation}
\label{Q0}
    Q_0{[\mathrm{barn}]} = \sqrt{ \frac{ 16 \pi B(E2)\uparrow {[e^2\,\mathrm{barn}^2]}}{5e^2} }. 
\end{equation}

As for the second part, we numerically solve Eq.~\eqref{H} within the dual-kinetic-balance approach \cite{PRL.93.130405} and calculate the ND corrections defined by Eqs.~(\ref{dE}, \ref{dg}).


\section{results and discussion}
\label{sec:discuss}

As mentioned in Sec.~\ref{sec:nuclear deformation corrections}, the nuclear $Q_0$ data used to calculate ND corrections are taken from Refs.~\cite{STONE2016, STONE2021} and Ref.~\cite{PRITYCHENKO2016}. 
The $Q$ values reported in Ref.~\cite{STONE2021} are experimentally measured by methods such as atomic spectroscopy, nuclear magnetic resonance, etc. 
Most of the $Q$ values have well-defined signs indicating the shape of the corresponding nuclear charge distributions, and for those which do not, we assume them to be positive. 
In addition, some nuclei in Ref.~\cite{STONE2021} do not have $Q$ data measured from nuclear ground states (277 nuclei). 
The corresponding ND corrections are then calculated using the $Q$ from the lowest available nuclear excited state. 
The $Q_0$ values obtained from the second data source \cite{PRITYCHENKO2016} are calculated from experimentally measured $B(E2)\uparrow$ transition rates, using Eq.~\eqref{Q0}, which implicitly assume positive $Q_0$. 
The data calculated from $Q_0$ with different signs and different references are separated on the plot by different marker styles (see Fig.~\ref{fig:1}).

In Fig.~\ref{fig:1} and Fig.~\ref{fig:2}, we present the ND corrections for $\delta E_{\mathrm{tran,ND}}$ and $\Delta g_{\mathrm{ND}}$, respectively, for 1155 nuclei. 
The overall trend is that all three ND corrections are positively correlated with $Z$ number. 
However, some individual ones appear to be significantly different compared with their neighbouring nuclei. 
For example, the nuclei on shell-closure feature smaller deformations, whereas the edges of the isotope sequences are more likely to possess a higher degree of shape distortion and the corresponding ND effects. 
More importantly, for the nuclei with negative $Q_0$ values, the ND corrections are systematically higher than for other similar nuclei, highlighting the critical importance of the measured $Q$ sign. 
Therefore, we conclude that the ND corrections based on the $Q$ values with unspecified signs are less reliable.

There are tens of special nuclei in Ref.~\cite{STONE2021} that have either two measured $Q$ values or two nuclear spins $I$, and correspondingly, two values for $\delta E_{\mathrm{ND}}$ and $\Delta g_{\mathrm{ND}}$, whereas Figs.~\ref{fig:1} and \ref{fig:2} only present one of each. 
For some of these nuclei, the differences between the two values are not compatible, in particular for the cases where the two $Q$ values have opposite signs, i.e.~$\prescript{94}{42}{\mathrm{Mo}}$, $\prescript{96}{42}{\mathrm{Mo}}$, $\prescript{134}{56}{\mathrm{Ba}}$, and $\prescript{136}{56}{\mathrm{Ba}}$.

The figures only show the results for $1s$ electronic state {or $1s_{1/2} \rightarrow 2p_{1/2}$ transition}, since the effect is maximized at the ground states. 
For excited states, our estimations show that ND corrections lose up to one order of magnitude by every principal, orbital momentum and total angular momentum quantum numbers.

Previous theoretical analyses \cite{PRL.108.063005, PRA.99.012505} of ND correction to the $g$ factor were based on strictly positive $Q_0$ values calculated from $B(E2)\uparrow$ transition rates with Eq.~\eqref{Q0}.
To avoid ambiguity in the sign of $Q_0$ we adopt the values in Ref.~\cite{STONE2021} instead, together with the measured signs, if available. 
This causes a slight discrepancy between our results and the corresponding data in Refs.~\cite{PRL.108.063005, PRA.99.012505}, which can be removed by using the same $\beta_2$ reported in their works \cite{PRL.108.063005, PRA.99.012505}.

\section{the search for new physics}
Recently, experimentally observed nonlinearity in the King plot of $\mathrm{Yb}^+$ ions \cite{PRL.125.123002, PRX.12.021033} led to a discussion on the existence of a new boson particle acting as a force carrier between electrons and neutrons. 
This is attributed to the observed nonlinearity being three times the experimental uncertainty, $3\sigma$ \cite{PRL.125.123002}. 
However, to claim the existence of new particles, one has to analyze nonlinearities from the existing physics, i.e.~quadratic field shifts and ND effects~\cite{PRA.103.L030801, PRL.125.123002}.

The five isotopes $^{168,170,172,174,176}\mathrm{Yb}^+$ and two transitions $6 \, ^{2}S_{1/2} \rightarrow 5 \, ^{2}D_{3/2} , \, 5 \, ^{2}D_{5/2}$ have been chosen to generate the King plot \cite{PRL.125.123002}.
Both transition energies have measurement precision of $\sim 300$~Hz. 
To estimate the ND energy corrections, we use perturbation theory with a perturbing potential $\Delta V = V^{(c\beta_2)} - V^{(c0)}$, defined by Eqs.~(\ref{V(r)}, \ref{dFermi}). 
The unperturbed many-electron wavefunctions of $\mathrm{Yb}^+$ are calculated using relativistic Hartree–Fock method, implemented with GRASP2018 package \cite{Fischer2019GRASP}. 
Since both transitions involve only one valence electron, we use
\begin{equation}
\label{eq:dE_diff}
    \Delta E^A_\mathrm{diff,ND} = \bra{\Psi^A_{d3}} \Delta V \ket{\Psi^A_{d3}} - \bra{\Psi^A_\mathrm{d5}} \Delta V \ket{\Psi^A_{d5}}
\end{equation}
to evaluate the ND correction to the difference between the two transition energies, where $\ket{\Psi^A_{d3}}$ and $\ket{\Psi^A_{d5}}$ are wavefunctions of the valence electron in $5d_{3/2}$ and $5d_{5/2}$ configurations for isotope $A$.

Unfortunately, the available $Q$ values measured from nuclear ground states only cover $^{158,160,162,164,166,168}\mathrm{Yb}^+$ isotopes \cite{PRITYCHENKO2016}. 
Among these isotopes, our calculations predict the ND effect monotonically increases with the mass number $A$ up to $\Delta E_\mathrm{diff,ND}^{168} = 312$~Hz. 
The increase between isotopes $( \Delta E_\mathrm{diff,ND}^{A+2}-\Delta E_\mathrm{diff,ND}^{A} )$ is $\sim 30$ to 50~Hz, which quantifies the King plot nonlinearity. 
Even though our results suggest that ND effects are not monotonic with respect to $A$ in general (see Fig.~\ref{fig:1}), it is believed that the five observed~\cite{PRL.125.123002} isotopes have the same nuclear ground state deformation $\beta_2 \approx 0.3$ \cite{PRA.103.L030801} as $^{166,168}\mathrm{Yb}^+$ examined above.  
Therefore we expect the ND contribution to the King plot nonlinearity 
to be at least on the same level of $\sim 30$ to 50~Hz. 
Summarizing, our estimation implies the ND effect alone is most probably not sufficient to explain the observed King plot nonlinearity~\cite{PRL.125.123002}. Further improvement  requires knowledge of the nuclear quadrupole moments for the nuclei under consideration, more accurate calculations with fully-correlated many-electron wavefunctions, and more realistic and sophisticated nuclear models.

\section{conclusion}
\label{sec:conclusion}
To conclude, by analyzing the available nuclear data over 1100 entries, we present the most complete picture of ND corrections for different observables.
We minimize the ambiguity in the signs of adopted intrinsic nuclear quadrupole moment $Q_0$ by carefully selecting the data sources. 
The small portion of nuclei with negative $Q_0$ reveal the importance of the oblate/prolate type of deformation, resulting in a difference of up to three orders of magnitude due to the sign flip. 
All calculated ND corrections are categorized based on the availability of the measured signs of $Q$ values. 
Our results show that the signs of $Q$ could be even more important than their absolute values for ND effects. 
Also, for the commonly accepted parameter $a$ in Fermi distribution, we modified the value for light nuclei to avoid it being unreasonably large.

Although nuclei with higher $Z$ numbers tend to be more deformed, the nuclear charts of deformation corrections also show significant non-monotonicity and case-by-case particularity. 
Finally, we can see that the current experiments are already approaching the level where ND effects are visible.
In such cases, customized and more sophisticated nuclear models as well as reliable data on nuclear parameters are required for a more accurate determination of the ND effects.


\acknowledgments

This article comprises parts of the PhD thesis work of Z. S. to be submitted to the Heidelberg University, Germany. 


\bibliography{ref}

\begin{thebibliography}{31}%
\makeatletter
\providecommand \@ifxundefined [1]{%
 \@ifx{#1\undefined}
}%
\providecommand \@ifnum [1]{%
 \ifnum #1\expandafter \@firstoftwo
 \else \expandafter \@secondoftwo
 \fi
}%
\providecommand \@ifx [1]{%
 \ifx #1\expandafter \@firstoftwo
 \else \expandafter \@secondoftwo
 \fi
}%
\providecommand \natexlab [1]{#1}%
\providecommand \enquote  [1]{``#1''}%
\providecommand \bibnamefont  [1]{#1}%
\providecommand \bibfnamefont [1]{#1}%
\providecommand \citenamefont [1]{#1}%
\providecommand \href@noop [0]{\@secondoftwo}%
\providecommand \href [0]{\begingroup \@sanitize@url \@href}%
\providecommand \@href[1]{\@@startlink{#1}\@@href}%
\providecommand \@@href[1]{\endgroup#1\@@endlink}%
\providecommand \@sanitize@url [0]{\catcode `\\12\catcode `\$12\catcode `\&12\catcode `\#12\catcode `\^12\catcode `\_12\catcode `\%12\relax}%
\providecommand \@@startlink[1]{}%
\providecommand \@@endlink[0]{}%
\providecommand \url  [0]{\begingroup\@sanitize@url \@url }%
\providecommand \@url [1]{\endgroup\@href {#1}{\urlprefix }}%
\providecommand \urlprefix  [0]{URL }%
\providecommand \Eprint [0]{\href }%
\providecommand \doibase [0]{https://doi.org/}%
\providecommand \selectlanguage [0]{\@gobble}%
\providecommand \bibinfo  [0]{\@secondoftwo}%
\providecommand \bibfield  [0]{\@secondoftwo}%
\providecommand \translation [1]{[#1]}%
\providecommand \BibitemOpen [0]{}%
\providecommand \bibitemStop [0]{}%
\providecommand \bibitemNoStop [0]{.\EOS\space}%
\providecommand \EOS [0]{\spacefactor3000\relax}%
\providecommand \BibitemShut  [1]{\csname bibitem#1\endcsname}%
\let\auto@bib@innerbib\@empty
\bibitem [{\citenamefont {Kozhedub}\ \emph {et~al.}(2008)\citenamefont {Kozhedub}, \citenamefont {Andreev}, \citenamefont {Shabaev}, \citenamefont {Tupitsyn}, \citenamefont {Brandau}, \citenamefont {Kozhuharov}, \citenamefont {Plunien},\ and\ \citenamefont {St\"ohlker}}]{PRA.77.032501}%
  \BibitemOpen
  \bibfield  {author} {\bibinfo {author} {\bibfnamefont {Y.~S.}\ \bibnamefont {Kozhedub}}, \bibinfo {author} {\bibfnamefont {O.~V.}\ \bibnamefont {Andreev}}, \bibinfo {author} {\bibfnamefont {V.~M.}\ \bibnamefont {Shabaev}}, \bibinfo {author} {\bibfnamefont {I.~I.}\ \bibnamefont {Tupitsyn}}, \bibinfo {author} {\bibfnamefont {C.}~\bibnamefont {Brandau}}, \bibinfo {author} {\bibfnamefont {C.}~\bibnamefont {Kozhuharov}}, \bibinfo {author} {\bibfnamefont {G.}~\bibnamefont {Plunien}},\ and\ \bibinfo {author} {\bibfnamefont {T.}~\bibnamefont {St\"ohlker}},\ }\href {https://doi.org/10.1103/PhysRevA.77.032501} {\bibfield  {journal} {\bibinfo  {journal} {Phys. Rev. A}\ }\textbf {\bibinfo {volume} {77}},\ \bibinfo {pages} {032501} (\bibinfo {year} {2008})}\BibitemShut {NoStop}%
\bibitem [{\citenamefont {Zatorski}\ \emph {et~al.}(2012)\citenamefont {Zatorski}, \citenamefont {Oreshkina}, \citenamefont {Keitel},\ and\ \citenamefont {Harman}}]{PRL.108.063005}%
  \BibitemOpen
  \bibfield  {author} {\bibinfo {author} {\bibfnamefont {J.}~\bibnamefont {Zatorski}}, \bibinfo {author} {\bibfnamefont {N.~S.}\ \bibnamefont {Oreshkina}}, \bibinfo {author} {\bibfnamefont {C.~H.}\ \bibnamefont {Keitel}},\ and\ \bibinfo {author} {\bibfnamefont {Z.}~\bibnamefont {Harman}},\ }\href {https://doi.org/10.1103/PhysRevLett.108.063005} {\bibfield  {journal} {\bibinfo  {journal} {Phys. Rev. Lett.}\ }\textbf {\bibinfo {volume} {108}},\ \bibinfo {pages} {063005} (\bibinfo {year} {2012})}\BibitemShut {NoStop}%
\bibitem [{\citenamefont {Michel}\ \emph {et~al.}(2019)\citenamefont {Michel}, \citenamefont {Zatorski}, \citenamefont {Oreshkina},\ and\ \citenamefont {Keitel}}]{PRA.99.012505}%
  \BibitemOpen
  \bibfield  {author} {\bibinfo {author} {\bibfnamefont {N.}~\bibnamefont {Michel}}, \bibinfo {author} {\bibfnamefont {J.}~\bibnamefont {Zatorski}}, \bibinfo {author} {\bibfnamefont {N.~S.}\ \bibnamefont {Oreshkina}},\ and\ \bibinfo {author} {\bibfnamefont {C.~H.}\ \bibnamefont {Keitel}},\ }\href {https://doi.org/10.1103/PhysRevA.99.012505} {\bibfield  {journal} {\bibinfo  {journal} {Phys. Rev. A}\ }\textbf {\bibinfo {volume} {99}},\ \bibinfo {pages} {012505} (\bibinfo {year} {2019})}\BibitemShut {NoStop}%
\bibitem [{\citenamefont {Dragani\ifmmode~\acute{c}\else \'{c}\fi{}}\ \emph {et~al.}(2003)\citenamefont {Dragani\ifmmode~\acute{c}\else \'{c}\fi{}}, \citenamefont {Crespo L\'opez-Urrutia}, \citenamefont {DuBois}, \citenamefont {Fritzsche}, \citenamefont {Shabaev}, \citenamefont {Orts}, \citenamefont {Tupitsyn}, \citenamefont {Zou},\ and\ \citenamefont {Ullrich}}]{PRL.91.183001}%
  \BibitemOpen
  \bibfield  {author} {\bibinfo {author} {\bibfnamefont {I.}~\bibnamefont {Dragani\ifmmode~\acute{c}\else \'{c}\fi{}}}, \bibinfo {author} {\bibfnamefont {J.~R.}\ \bibnamefont {Crespo L\'opez-Urrutia}}, \bibinfo {author} {\bibfnamefont {R.}~\bibnamefont {DuBois}}, \bibinfo {author} {\bibfnamefont {S.}~\bibnamefont {Fritzsche}}, \bibinfo {author} {\bibfnamefont {V.~M.}\ \bibnamefont {Shabaev}}, \bibinfo {author} {\bibfnamefont {R.~S.}\ \bibnamefont {Orts}}, \bibinfo {author} {\bibfnamefont {I.~I.}\ \bibnamefont {Tupitsyn}}, \bibinfo {author} {\bibfnamefont {Y.}~\bibnamefont {Zou}},\ and\ \bibinfo {author} {\bibfnamefont {J.}~\bibnamefont {Ullrich}},\ }\href {https://doi.org/10.1103/PhysRevLett.91.183001} {\bibfield  {journal} {\bibinfo  {journal} {Phys. Rev. Lett.}\ }\textbf {\bibinfo {volume} {91}},\ \bibinfo {pages} {183001} (\bibinfo {year} {2003})}\BibitemShut {NoStop}%
\bibitem [{\citenamefont {Gumberidze}\ \emph {et~al.}(2005)\citenamefont {Gumberidze}, \citenamefont {St\"ohlker}, \citenamefont {Bana\ifmmode~\acute{s}\else \'{s}\fi{}}, \citenamefont {Beckert}, \citenamefont {Beller}, \citenamefont {Beyer}, \citenamefont {Bosch}, \citenamefont {Hagmann}, \citenamefont {Kozhuharov}, \citenamefont {Liesen}, \citenamefont {Nolden}, \citenamefont {Ma}, \citenamefont {Mokler}, \citenamefont {Steck}, \citenamefont {Sierpowski},\ and\ \citenamefont {Tashenov}}]{PRL.94.223001}%
  \BibitemOpen
  \bibfield  {author} {\bibinfo {author} {\bibfnamefont {A.}~\bibnamefont {Gumberidze}}, \bibinfo {author} {\bibfnamefont {T.}~\bibnamefont {St\"ohlker}}, \bibinfo {author} {\bibfnamefont {D.}~\bibnamefont {Bana\ifmmode~\acute{s}\else \'{s}\fi{}}}, \bibinfo {author} {\bibfnamefont {K.}~\bibnamefont {Beckert}}, \bibinfo {author} {\bibfnamefont {P.}~\bibnamefont {Beller}}, \bibinfo {author} {\bibfnamefont {H.~F.}\ \bibnamefont {Beyer}}, \bibinfo {author} {\bibfnamefont {F.}~\bibnamefont {Bosch}}, \bibinfo {author} {\bibfnamefont {S.}~\bibnamefont {Hagmann}}, \bibinfo {author} {\bibfnamefont {C.}~\bibnamefont {Kozhuharov}}, \bibinfo {author} {\bibfnamefont {D.}~\bibnamefont {Liesen}}, \bibinfo {author} {\bibfnamefont {F.}~\bibnamefont {Nolden}}, \bibinfo {author} {\bibfnamefont {X.}~\bibnamefont {Ma}}, \bibinfo {author} {\bibfnamefont {P.~H.}\ \bibnamefont {Mokler}}, \bibinfo {author} {\bibfnamefont {M.}~\bibnamefont {Steck}}, \bibinfo {author} {\bibfnamefont {D.}~\bibnamefont {Sierpowski}},\ and\ \bibinfo
  {author} {\bibfnamefont {S.}~\bibnamefont {Tashenov}},\ }\href {https://doi.org/10.1103/PhysRevLett.94.223001} {\bibfield  {journal} {\bibinfo  {journal} {Phys. Rev. Lett.}\ }\textbf {\bibinfo {volume} {94}},\ \bibinfo {pages} {223001} (\bibinfo {year} {2005})}\BibitemShut {NoStop}%
\bibitem [{\citenamefont {Gunst}\ \emph {et~al.}(2013)\citenamefont {Gunst}, \citenamefont {Surzhykov}, \citenamefont {Artemyev}, \citenamefont {Fritzsche}, \citenamefont {Tashenov}, \citenamefont {Maiorova}, \citenamefont {Shabaev},\ and\ \citenamefont {St\"ohlker}}]{PRA.87.032714}%
  \BibitemOpen
  \bibfield  {author} {\bibinfo {author} {\bibfnamefont {J.}~\bibnamefont {Gunst}}, \bibinfo {author} {\bibfnamefont {A.}~\bibnamefont {Surzhykov}}, \bibinfo {author} {\bibfnamefont {A.}~\bibnamefont {Artemyev}}, \bibinfo {author} {\bibfnamefont {S.}~\bibnamefont {Fritzsche}}, \bibinfo {author} {\bibfnamefont {S.}~\bibnamefont {Tashenov}}, \bibinfo {author} {\bibfnamefont {A.}~\bibnamefont {Maiorova}}, \bibinfo {author} {\bibfnamefont {V.~M.}\ \bibnamefont {Shabaev}},\ and\ \bibinfo {author} {\bibfnamefont {T.}~\bibnamefont {St\"ohlker}},\ }\href {https://doi.org/10.1103/PhysRevA.87.032714} {\bibfield  {journal} {\bibinfo  {journal} {Phys. Rev. A}\ }\textbf {\bibinfo {volume} {87}},\ \bibinfo {pages} {032714} (\bibinfo {year} {2013})}\BibitemShut {NoStop}%
\bibitem [{\citenamefont {J\"org}\ \emph {et~al.}(2015)\citenamefont {J\"org}, \citenamefont {Hu}, \citenamefont {Bekker}, \citenamefont {Blessenohl}, \citenamefont {Hollain}, \citenamefont {Fritzsche}, \citenamefont {Surzhykov}, \citenamefont {Crespo L\'opez-Urrutia},\ and\ \citenamefont {Tashenov}}]{PRA.91.042705}%
  \BibitemOpen
  \bibfield  {author} {\bibinfo {author} {\bibfnamefont {H.}~\bibnamefont {J\"org}}, \bibinfo {author} {\bibfnamefont {Z.}~\bibnamefont {Hu}}, \bibinfo {author} {\bibfnamefont {H.}~\bibnamefont {Bekker}}, \bibinfo {author} {\bibfnamefont {M.~A.}\ \bibnamefont {Blessenohl}}, \bibinfo {author} {\bibfnamefont {D.}~\bibnamefont {Hollain}}, \bibinfo {author} {\bibfnamefont {S.}~\bibnamefont {Fritzsche}}, \bibinfo {author} {\bibfnamefont {A.}~\bibnamefont {Surzhykov}}, \bibinfo {author} {\bibfnamefont {J.~R.}\ \bibnamefont {Crespo L\'opez-Urrutia}},\ and\ \bibinfo {author} {\bibfnamefont {S.}~\bibnamefont {Tashenov}},\ }\href {https://doi.org/10.1103/PhysRevA.91.042705} {\bibfield  {journal} {\bibinfo  {journal} {Phys. Rev. A}\ }\textbf {\bibinfo {volume} {91}},\ \bibinfo {pages} {042705} (\bibinfo {year} {2015})}\BibitemShut {NoStop}%
\bibitem [{\citenamefont {Heavner}\ \emph {et~al.}(2014)\citenamefont {Heavner}, \citenamefont {Donley}, \citenamefont {Levi}, \citenamefont {Costanzo}, \citenamefont {Parker}, \citenamefont {Shirley}, \citenamefont {Ashby}, \citenamefont {Barlow},\ and\ \citenamefont {Jefferts}}]{Heavner_2014}%
  \BibitemOpen
  \bibfield  {author} {\bibinfo {author} {\bibfnamefont {T.~P.}\ \bibnamefont {Heavner}}, \bibinfo {author} {\bibfnamefont {E.~A.}\ \bibnamefont {Donley}}, \bibinfo {author} {\bibfnamefont {F.}~\bibnamefont {Levi}}, \bibinfo {author} {\bibfnamefont {G.}~\bibnamefont {Costanzo}}, \bibinfo {author} {\bibfnamefont {T.~E.}\ \bibnamefont {Parker}}, \bibinfo {author} {\bibfnamefont {J.~H.}\ \bibnamefont {Shirley}}, \bibinfo {author} {\bibfnamefont {N.}~\bibnamefont {Ashby}}, \bibinfo {author} {\bibfnamefont {S.}~\bibnamefont {Barlow}},\ and\ \bibinfo {author} {\bibfnamefont {S.~R.}\ \bibnamefont {Jefferts}},\ }\href {https://doi.org/10.1088/0026-1394/51/3/174} {\bibfield  {journal} {\bibinfo  {journal} {Metrologia}\ }\textbf {\bibinfo {volume} {51}},\ \bibinfo {pages} {174} (\bibinfo {year} {2014})}\BibitemShut {NoStop}%
\bibitem [{\citenamefont {Odom}\ \emph {et~al.}(2006)\citenamefont {Odom}, \citenamefont {Hanneke}, \citenamefont {D'Urso},\ and\ \citenamefont {Gabrielse}}]{PRL.97.030801}%
  \BibitemOpen
  \bibfield  {author} {\bibinfo {author} {\bibfnamefont {B.}~\bibnamefont {Odom}}, \bibinfo {author} {\bibfnamefont {D.}~\bibnamefont {Hanneke}}, \bibinfo {author} {\bibfnamefont {B.}~\bibnamefont {D'Urso}},\ and\ \bibinfo {author} {\bibfnamefont {G.}~\bibnamefont {Gabrielse}},\ }\href {https://doi.org/10.1103/PhysRevLett.97.030801} {\bibfield  {journal} {\bibinfo  {journal} {Phys. Rev. Lett.}\ }\textbf {\bibinfo {volume} {97}},\ \bibinfo {pages} {030801} (\bibinfo {year} {2006})}\BibitemShut {NoStop}%
\bibitem [{\citenamefont {H\"affner}\ \emph {et~al.}(2000)\citenamefont {H\"affner}, \citenamefont {Beier}, \citenamefont {Hermanspahn}, \citenamefont {Kluge}, \citenamefont {Quint}, \citenamefont {Stahl}, \citenamefont {Verd\'u},\ and\ \citenamefont {Werth}}]{PRL.85.5308}%
  \BibitemOpen
  \bibfield  {author} {\bibinfo {author} {\bibfnamefont {H.}~\bibnamefont {H\"affner}}, \bibinfo {author} {\bibfnamefont {T.}~\bibnamefont {Beier}}, \bibinfo {author} {\bibfnamefont {N.}~\bibnamefont {Hermanspahn}}, \bibinfo {author} {\bibfnamefont {H.-J.}\ \bibnamefont {Kluge}}, \bibinfo {author} {\bibfnamefont {W.}~\bibnamefont {Quint}}, \bibinfo {author} {\bibfnamefont {S.}~\bibnamefont {Stahl}}, \bibinfo {author} {\bibfnamefont {J.}~\bibnamefont {Verd\'u}},\ and\ \bibinfo {author} {\bibfnamefont {G.}~\bibnamefont {Werth}},\ }\href {https://doi.org/10.1103/PhysRevLett.85.5308} {\bibfield  {journal} {\bibinfo  {journal} {Phys. Rev. Lett.}\ }\textbf {\bibinfo {volume} {85}},\ \bibinfo {pages} {5308} (\bibinfo {year} {2000})}\BibitemShut {NoStop}%
\bibitem [{\citenamefont {Verd\'u}\ \emph {et~al.}(2004)\citenamefont {Verd\'u}, \citenamefont {Djeki\ifmmode~\acute{c}\else \'{c}\fi{}}, \citenamefont {Stahl}, \citenamefont {Valenzuela}, \citenamefont {Vogel}, \citenamefont {Werth}, \citenamefont {Beier}, \citenamefont {Kluge},\ and\ \citenamefont {Quint}}]{PRL.92.093002}%
  \BibitemOpen
  \bibfield  {author} {\bibinfo {author} {\bibfnamefont {J.}~\bibnamefont {Verd\'u}}, \bibinfo {author} {\bibfnamefont {S.}~\bibnamefont {Djeki\ifmmode~\acute{c}\else \'{c}\fi{}}}, \bibinfo {author} {\bibfnamefont {S.}~\bibnamefont {Stahl}}, \bibinfo {author} {\bibfnamefont {T.}~\bibnamefont {Valenzuela}}, \bibinfo {author} {\bibfnamefont {M.}~\bibnamefont {Vogel}}, \bibinfo {author} {\bibfnamefont {G.}~\bibnamefont {Werth}}, \bibinfo {author} {\bibfnamefont {T.}~\bibnamefont {Beier}}, \bibinfo {author} {\bibfnamefont {H.-J.}\ \bibnamefont {Kluge}},\ and\ \bibinfo {author} {\bibfnamefont {W.}~\bibnamefont {Quint}},\ }\href {https://doi.org/10.1103/PhysRevLett.92.093002} {\bibfield  {journal} {\bibinfo  {journal} {Phys. Rev. Lett.}\ }\textbf {\bibinfo {volume} {92}},\ \bibinfo {pages} {093002} (\bibinfo {year} {2004})}\BibitemShut {NoStop}%
\bibitem [{\citenamefont {Sturm}\ \emph {et~al.}(2011)\citenamefont {Sturm}, \citenamefont {Wagner}, \citenamefont {Schabinger}, \citenamefont {Zatorski}, \citenamefont {Harman}, \citenamefont {Quint}, \citenamefont {Werth}, \citenamefont {Keitel},\ and\ \citenamefont {Blaum}}]{PRL.107.023002}%
  \BibitemOpen
  \bibfield  {author} {\bibinfo {author} {\bibfnamefont {S.}~\bibnamefont {Sturm}}, \bibinfo {author} {\bibfnamefont {A.}~\bibnamefont {Wagner}}, \bibinfo {author} {\bibfnamefont {B.}~\bibnamefont {Schabinger}}, \bibinfo {author} {\bibfnamefont {J.}~\bibnamefont {Zatorski}}, \bibinfo {author} {\bibfnamefont {Z.}~\bibnamefont {Harman}}, \bibinfo {author} {\bibfnamefont {W.}~\bibnamefont {Quint}}, \bibinfo {author} {\bibfnamefont {G.}~\bibnamefont {Werth}}, \bibinfo {author} {\bibfnamefont {C.~H.}\ \bibnamefont {Keitel}},\ and\ \bibinfo {author} {\bibfnamefont {K.}~\bibnamefont {Blaum}},\ }\href {https://doi.org/10.1103/PhysRevLett.107.023002} {\bibfield  {journal} {\bibinfo  {journal} {Phys. Rev. Lett.}\ }\textbf {\bibinfo {volume} {107}},\ \bibinfo {pages} {023002} (\bibinfo {year} {2011})}\BibitemShut {NoStop}%
\bibitem [{\citenamefont {Arapoglou}\ \emph {et~al.}(2019)\citenamefont {Arapoglou}, \citenamefont {Egl}, \citenamefont {H\"ocker}, \citenamefont {Sailer}, \citenamefont {Tu}, \citenamefont {Weigel}, \citenamefont {Wolf}, \citenamefont {Cakir}, \citenamefont {Yerokhin}, \citenamefont {Oreshkina}, \citenamefont {Agababaev}, \citenamefont {Volotka}, \citenamefont {Zinenko}, \citenamefont {Glazov}, \citenamefont {Harman}, \citenamefont {Keitel}, \citenamefont {Sturm},\ and\ \citenamefont {Blaum}}]{PRL.122.253001}%
  \BibitemOpen
  \bibfield  {author} {\bibinfo {author} {\bibfnamefont {I.}~\bibnamefont {Arapoglou}}, \bibinfo {author} {\bibfnamefont {A.}~\bibnamefont {Egl}}, \bibinfo {author} {\bibfnamefont {M.}~\bibnamefont {H\"ocker}}, \bibinfo {author} {\bibfnamefont {T.}~\bibnamefont {Sailer}}, \bibinfo {author} {\bibfnamefont {B.}~\bibnamefont {Tu}}, \bibinfo {author} {\bibfnamefont {A.}~\bibnamefont {Weigel}}, \bibinfo {author} {\bibfnamefont {R.}~\bibnamefont {Wolf}}, \bibinfo {author} {\bibfnamefont {H.}~\bibnamefont {Cakir}}, \bibinfo {author} {\bibfnamefont {V.~A.}\ \bibnamefont {Yerokhin}}, \bibinfo {author} {\bibfnamefont {N.~S.}\ \bibnamefont {Oreshkina}}, \bibinfo {author} {\bibfnamefont {V.~A.}\ \bibnamefont {Agababaev}}, \bibinfo {author} {\bibfnamefont {A.~V.}\ \bibnamefont {Volotka}}, \bibinfo {author} {\bibfnamefont {D.~V.}\ \bibnamefont {Zinenko}}, \bibinfo {author} {\bibfnamefont {D.~A.}\ \bibnamefont {Glazov}}, \bibinfo {author} {\bibfnamefont {Z.}~\bibnamefont {Harman}}, \bibinfo {author} {\bibfnamefont {C.~H.}\
  \bibnamefont {Keitel}}, \bibinfo {author} {\bibfnamefont {S.}~\bibnamefont {Sturm}},\ and\ \bibinfo {author} {\bibfnamefont {K.}~\bibnamefont {Blaum}},\ }\href {https://doi.org/10.1103/PhysRevLett.122.253001} {\bibfield  {journal} {\bibinfo  {journal} {Phys. Rev. Lett.}\ }\textbf {\bibinfo {volume} {122}},\ \bibinfo {pages} {253001} (\bibinfo {year} {2019})}\BibitemShut {NoStop}%
\bibitem [{\citenamefont {Glazov}\ \emph {et~al.}(2019)\citenamefont {Glazov}, \citenamefont {K\"ohler-Langes}, \citenamefont {Volotka}, \citenamefont {Blaum}, \citenamefont {Hei\ss{}e}, \citenamefont {Plunien}, \citenamefont {Quint}, \citenamefont {Rau}, \citenamefont {Shabaev}, \citenamefont {Sturm},\ and\ \citenamefont {Werth}}]{PRL.123.173001}%
  \BibitemOpen
  \bibfield  {author} {\bibinfo {author} {\bibfnamefont {D.~A.}\ \bibnamefont {Glazov}}, \bibinfo {author} {\bibfnamefont {F.}~\bibnamefont {K\"ohler-Langes}}, \bibinfo {author} {\bibfnamefont {A.~V.}\ \bibnamefont {Volotka}}, \bibinfo {author} {\bibfnamefont {K.}~\bibnamefont {Blaum}}, \bibinfo {author} {\bibfnamefont {F.}~\bibnamefont {Hei\ss{}e}}, \bibinfo {author} {\bibfnamefont {G.}~\bibnamefont {Plunien}}, \bibinfo {author} {\bibfnamefont {W.}~\bibnamefont {Quint}}, \bibinfo {author} {\bibfnamefont {S.}~\bibnamefont {Rau}}, \bibinfo {author} {\bibfnamefont {V.~M.}\ \bibnamefont {Shabaev}}, \bibinfo {author} {\bibfnamefont {S.}~\bibnamefont {Sturm}},\ and\ \bibinfo {author} {\bibfnamefont {G.}~\bibnamefont {Werth}},\ }\href {https://doi.org/10.1103/PhysRevLett.123.173001} {\bibfield  {journal} {\bibinfo  {journal} {Phys. Rev. Lett.}\ }\textbf {\bibinfo {volume} {123}},\ \bibinfo {pages} {173001} (\bibinfo {year} {2019})}\BibitemShut {NoStop}%
\bibitem [{\citenamefont {Debierre}\ \emph {et~al.}(2020)\citenamefont {Debierre}, \citenamefont {Keitel},\ and\ \citenamefont {Harman}}]{DEBIERRE2020135527}%
  \BibitemOpen
  \bibfield  {author} {\bibinfo {author} {\bibfnamefont {V.}~\bibnamefont {Debierre}}, \bibinfo {author} {\bibfnamefont {C.}~\bibnamefont {Keitel}},\ and\ \bibinfo {author} {\bibfnamefont {Z.}~\bibnamefont {Harman}},\ }\href {https://doi.org/https://doi.org/10.1016/j.physletb.2020.135527} {\bibfield  {journal} {\bibinfo  {journal} {Physics Letters B}\ }\textbf {\bibinfo {volume} {807}},\ \bibinfo {pages} {135527} (\bibinfo {year} {2020})}\BibitemShut {NoStop}%
\bibitem [{\citenamefont {Debierre}\ \emph {et~al.}(2022)\citenamefont {Debierre}, \citenamefont {Oreshkina}, \citenamefont {Valuev}, \citenamefont {Harman},\ and\ \citenamefont {Keitel}}]{PRA.106.062801}%
  \BibitemOpen
  \bibfield  {author} {\bibinfo {author} {\bibfnamefont {V.}~\bibnamefont {Debierre}}, \bibinfo {author} {\bibfnamefont {N.~S.}\ \bibnamefont {Oreshkina}}, \bibinfo {author} {\bibfnamefont {I.~A.}\ \bibnamefont {Valuev}}, \bibinfo {author} {\bibfnamefont {Z.}~\bibnamefont {Harman}},\ and\ \bibinfo {author} {\bibfnamefont {C.~H.}\ \bibnamefont {Keitel}},\ }\href {https://doi.org/10.1103/PhysRevA.106.062801} {\bibfield  {journal} {\bibinfo  {journal} {Phys. Rev. A}\ }\textbf {\bibinfo {volume} {106}},\ \bibinfo {pages} {062801} (\bibinfo {year} {2022})}\BibitemShut {NoStop}%
\bibitem [{\citenamefont {Seltzer}(1969)}]{PhysRev.188.1916}%
  \BibitemOpen
  \bibfield  {author} {\bibinfo {author} {\bibfnamefont {E.~C.}\ \bibnamefont {Seltzer}},\ }\href {https://doi.org/10.1103/PhysRev.188.1916} {\bibfield  {journal} {\bibinfo  {journal} {Phys. Rev.}\ }\textbf {\bibinfo {volume} {188}},\ \bibinfo {pages} {1916} (\bibinfo {year} {1969})}\BibitemShut {NoStop}%
\bibitem [{\citenamefont {Fricke}\ \emph {et~al.}(1995)\citenamefont {Fricke}, \citenamefont {Bernhardt}, \citenamefont {Heilig}, \citenamefont {Schaller}, \citenamefont {Schellenberg}, \citenamefont {Shera},\ and\ \citenamefont {Dejager}}]{FRICKE1995177}%
  \BibitemOpen
  \bibfield  {author} {\bibinfo {author} {\bibfnamefont {G.}~\bibnamefont {Fricke}}, \bibinfo {author} {\bibfnamefont {C.}~\bibnamefont {Bernhardt}}, \bibinfo {author} {\bibfnamefont {K.}~\bibnamefont {Heilig}}, \bibinfo {author} {\bibfnamefont {L.}~\bibnamefont {Schaller}}, \bibinfo {author} {\bibfnamefont {L.}~\bibnamefont {Schellenberg}}, \bibinfo {author} {\bibfnamefont {E.}~\bibnamefont {Shera}},\ and\ \bibinfo {author} {\bibfnamefont {C.}~\bibnamefont {Dejager}},\ }\href {https://doi.org/https://doi.org/10.1006/adnd.1995.1007} {\bibfield  {journal} {\bibinfo  {journal} {Atomic Data and Nuclear Data Tables}\ }\textbf {\bibinfo {volume} {60}},\ \bibinfo {pages} {177} (\bibinfo {year} {1995})}\BibitemShut {NoStop}%
\bibitem [{\citenamefont {Angeli}\ and\ \citenamefont {Marinova}(2013)}]{ANGELI201369}%
  \BibitemOpen
  \bibfield  {author} {\bibinfo {author} {\bibfnamefont {I.}~\bibnamefont {Angeli}}\ and\ \bibinfo {author} {\bibfnamefont {K.}~\bibnamefont {Marinova}},\ }\href {https://doi.org/https://doi.org/10.1016/j.adt.2011.12.006} {\bibfield  {journal} {\bibinfo  {journal} {Atomic Data and Nuclear Data Tables}\ }\textbf {\bibinfo {volume} {99}},\ \bibinfo {pages} {69} (\bibinfo {year} {2013})}\BibitemShut {NoStop}%
\bibitem [{\citenamefont {Delaunay}\ \emph {et~al.}(2017)\citenamefont {Delaunay}, \citenamefont {Ozeri}, \citenamefont {Perez},\ and\ \citenamefont {Soreq}}]{PRD.96.093001}%
  \BibitemOpen
  \bibfield  {author} {\bibinfo {author} {\bibfnamefont {C.}~\bibnamefont {Delaunay}}, \bibinfo {author} {\bibfnamefont {R.}~\bibnamefont {Ozeri}}, \bibinfo {author} {\bibfnamefont {G.}~\bibnamefont {Perez}},\ and\ \bibinfo {author} {\bibfnamefont {Y.}~\bibnamefont {Soreq}},\ }\href {https://doi.org/10.1103/PhysRevD.96.093001} {\bibfield  {journal} {\bibinfo  {journal} {Phys. Rev. D}\ }\textbf {\bibinfo {volume} {96}},\ \bibinfo {pages} {093001} (\bibinfo {year} {2017})}\BibitemShut {NoStop}%
\bibitem [{\citenamefont {Berengut}\ \emph {et~al.}(2018)\citenamefont {Berengut}, \citenamefont {Budker}, \citenamefont {Delaunay}, \citenamefont {Flambaum}, \citenamefont {Frugiuele}, \citenamefont {Fuchs}, \citenamefont {Grojean}, \citenamefont {Harnik}, \citenamefont {Ozeri}, \citenamefont {Perez},\ and\ \citenamefont {Soreq}}]{PRL.120.091801}%
  \BibitemOpen
  \bibfield  {author} {\bibinfo {author} {\bibfnamefont {J.~C.}\ \bibnamefont {Berengut}}, \bibinfo {author} {\bibfnamefont {D.}~\bibnamefont {Budker}}, \bibinfo {author} {\bibfnamefont {C.}~\bibnamefont {Delaunay}}, \bibinfo {author} {\bibfnamefont {V.~V.}\ \bibnamefont {Flambaum}}, \bibinfo {author} {\bibfnamefont {C.}~\bibnamefont {Frugiuele}}, \bibinfo {author} {\bibfnamefont {E.}~\bibnamefont {Fuchs}}, \bibinfo {author} {\bibfnamefont {C.}~\bibnamefont {Grojean}}, \bibinfo {author} {\bibfnamefont {R.}~\bibnamefont {Harnik}}, \bibinfo {author} {\bibfnamefont {R.}~\bibnamefont {Ozeri}}, \bibinfo {author} {\bibfnamefont {G.}~\bibnamefont {Perez}},\ and\ \bibinfo {author} {\bibfnamefont {Y.}~\bibnamefont {Soreq}},\ }\href {https://doi.org/10.1103/PhysRevLett.120.091801} {\bibfield  {journal} {\bibinfo  {journal} {Phys. Rev. Lett.}\ }\textbf {\bibinfo {volume} {120}},\ \bibinfo {pages} {091801} (\bibinfo {year} {2018})}\BibitemShut {NoStop}%
\bibitem [{\citenamefont {Allehabi}\ \emph {et~al.}(2021)\citenamefont {Allehabi}, \citenamefont {Dzuba}, \citenamefont {Flambaum},\ and\ \citenamefont {Afanasjev}}]{PRA.103.L030801}%
  \BibitemOpen
  \bibfield  {author} {\bibinfo {author} {\bibfnamefont {S.~O.}\ \bibnamefont {Allehabi}}, \bibinfo {author} {\bibfnamefont {V.~A.}\ \bibnamefont {Dzuba}}, \bibinfo {author} {\bibfnamefont {V.~V.}\ \bibnamefont {Flambaum}},\ and\ \bibinfo {author} {\bibfnamefont {A.~V.}\ \bibnamefont {Afanasjev}},\ }\href {https://doi.org/10.1103/PhysRevA.103.L030801} {\bibfield  {journal} {\bibinfo  {journal} {Phys. Rev. A}\ }\textbf {\bibinfo {volume} {103}},\ \bibinfo {pages} {L030801} (\bibinfo {year} {2021})}\BibitemShut {NoStop}%
\bibitem [{\citenamefont {Counts}\ \emph {et~al.}(2020)\citenamefont {Counts}, \citenamefont {Hur}, \citenamefont {Aude~Craik}, \citenamefont {Jeon}, \citenamefont {Leung}, \citenamefont {Berengut}, \citenamefont {Geddes}, \citenamefont {Kawasaki}, \citenamefont {Jhe},\ and\ \citenamefont {Vuleti\ifmmode~\acute{c}\else \'{c}\fi{}}}]{PRL.125.123002}%
  \BibitemOpen
  \bibfield  {author} {\bibinfo {author} {\bibfnamefont {I.}~\bibnamefont {Counts}}, \bibinfo {author} {\bibfnamefont {J.}~\bibnamefont {Hur}}, \bibinfo {author} {\bibfnamefont {D.~P.~L.}\ \bibnamefont {Aude~Craik}}, \bibinfo {author} {\bibfnamefont {H.}~\bibnamefont {Jeon}}, \bibinfo {author} {\bibfnamefont {C.}~\bibnamefont {Leung}}, \bibinfo {author} {\bibfnamefont {J.~C.}\ \bibnamefont {Berengut}}, \bibinfo {author} {\bibfnamefont {A.}~\bibnamefont {Geddes}}, \bibinfo {author} {\bibfnamefont {A.}~\bibnamefont {Kawasaki}}, \bibinfo {author} {\bibfnamefont {W.}~\bibnamefont {Jhe}},\ and\ \bibinfo {author} {\bibfnamefont {V.}~\bibnamefont {Vuleti\ifmmode~\acute{c}\else \'{c}\fi{}}},\ }\href {https://doi.org/10.1103/PhysRevLett.125.123002} {\bibfield  {journal} {\bibinfo  {journal} {Phys. Rev. Lett.}\ }\textbf {\bibinfo {volume} {125}},\ \bibinfo {pages} {123002} (\bibinfo {year} {2020})}\BibitemShut {NoStop}%
\bibitem [{\citenamefont {Ono}\ \emph {et~al.}(2022)\citenamefont {Ono}, \citenamefont {Saito}, \citenamefont {Ishiyama}, \citenamefont {Higomoto}, \citenamefont {Takano}, \citenamefont {Takasu}, \citenamefont {Yamamoto}, \citenamefont {Tanaka},\ and\ \citenamefont {Takahashi}}]{PRX.12.021033}%
  \BibitemOpen
  \bibfield  {author} {\bibinfo {author} {\bibfnamefont {K.}~\bibnamefont {Ono}}, \bibinfo {author} {\bibfnamefont {Y.}~\bibnamefont {Saito}}, \bibinfo {author} {\bibfnamefont {T.}~\bibnamefont {Ishiyama}}, \bibinfo {author} {\bibfnamefont {T.}~\bibnamefont {Higomoto}}, \bibinfo {author} {\bibfnamefont {T.}~\bibnamefont {Takano}}, \bibinfo {author} {\bibfnamefont {Y.}~\bibnamefont {Takasu}}, \bibinfo {author} {\bibfnamefont {Y.}~\bibnamefont {Yamamoto}}, \bibinfo {author} {\bibfnamefont {M.}~\bibnamefont {Tanaka}},\ and\ \bibinfo {author} {\bibfnamefont {Y.}~\bibnamefont {Takahashi}},\ }\href {https://doi.org/10.1103/PhysRevX.12.021033} {\bibfield  {journal} {\bibinfo  {journal} {Phys. Rev. X}\ }\textbf {\bibinfo {volume} {12}},\ \bibinfo {pages} {021033} (\bibinfo {year} {2022})}\BibitemShut {NoStop}%
\bibitem [{\citenamefont {Parpia}\ and\ \citenamefont {Mohanty}(1992)}]{PRA.46.3735}%
  \BibitemOpen
  \bibfield  {author} {\bibinfo {author} {\bibfnamefont {F.~A.}\ \bibnamefont {Parpia}}\ and\ \bibinfo {author} {\bibfnamefont {A.~K.}\ \bibnamefont {Mohanty}},\ }\href {https://doi.org/10.1103/PhysRevA.46.3735} {\bibfield  {journal} {\bibinfo  {journal} {Phys. Rev. A}\ }\textbf {\bibinfo {volume} {46}},\ \bibinfo {pages} {3735} (\bibinfo {year} {1992})}\BibitemShut {NoStop}%
\bibitem [{sup()}]{supplement}%
  \BibitemOpen
  \href@noop {} {\bibinfo {title} {See {S}upplemental {M}aterial at http://... for {ND} corrections to ground state eigenenergy and hyperfine splitting constant, as well as more numerical details.}}\BibitemShut {Stop}%
\bibitem [{\citenamefont {Stone}(2016)}]{STONE2016}%
  \BibitemOpen
  \bibfield  {author} {\bibinfo {author} {\bibfnamefont {N.}~\bibnamefont {Stone}},\ }\href {https://doi.org/https://doi.org/10.1016/j.adt.2015.12.002} {\bibfield  {journal} {\bibinfo  {journal} {Atomic Data and Nuclear Data Tables}\ }\textbf {\bibinfo {volume} {111-112}},\ \bibinfo {pages} {1} (\bibinfo {year} {2016})}\BibitemShut {NoStop}%
\bibitem [{\citenamefont {Stone}(2021)}]{STONE2021}%
  \BibitemOpen
  \bibfield  {author} {\bibinfo {author} {\bibfnamefont {N.}~\bibnamefont {Stone}},\ }\href {https://www-nds.iaea.org/publications/indc/indc-nds-0833.pdf} {\bibinfo {title} {Table of nuclear electric quadrupole moments}} (\bibinfo {year} {2021})\BibitemShut {NoStop}%
\bibitem [{\citenamefont {Pritychenko}\ \emph {et~al.}(2016)\citenamefont {Pritychenko}, \citenamefont {Birch}, \citenamefont {Singh},\ and\ \citenamefont {Horoi}}]{PRITYCHENKO2016}%
  \BibitemOpen
  \bibfield  {author} {\bibinfo {author} {\bibfnamefont {B.}~\bibnamefont {Pritychenko}}, \bibinfo {author} {\bibfnamefont {M.}~\bibnamefont {Birch}}, \bibinfo {author} {\bibfnamefont {B.}~\bibnamefont {Singh}},\ and\ \bibinfo {author} {\bibfnamefont {M.}~\bibnamefont {Horoi}},\ }\href {https://doi.org/https://doi.org/10.1016/j.adt.2015.10.001} {\bibfield  {journal} {\bibinfo  {journal} {Atomic Data and Nuclear Data Tables}\ }\textbf {\bibinfo {volume} {107}},\ \bibinfo {pages} {1} (\bibinfo {year} {2016})}\BibitemShut {NoStop}%
\bibitem [{\citenamefont {Shabaev}\ \emph {et~al.}(2004)\citenamefont {Shabaev}, \citenamefont {Tupitsyn}, \citenamefont {Yerokhin}, \citenamefont {Plunien},\ and\ \citenamefont {Soff}}]{PRL.93.130405}%
  \BibitemOpen
  \bibfield  {author} {\bibinfo {author} {\bibfnamefont {V.~M.}\ \bibnamefont {Shabaev}}, \bibinfo {author} {\bibfnamefont {I.~I.}\ \bibnamefont {Tupitsyn}}, \bibinfo {author} {\bibfnamefont {V.~A.}\ \bibnamefont {Yerokhin}}, \bibinfo {author} {\bibfnamefont {G.}~\bibnamefont {Plunien}},\ and\ \bibinfo {author} {\bibfnamefont {G.}~\bibnamefont {Soff}},\ }\href {https://doi.org/10.1103/PhysRevLett.93.130405} {\bibfield  {journal} {\bibinfo  {journal} {Phys. Rev. Lett.}\ }\textbf {\bibinfo {volume} {93}},\ \bibinfo {pages} {130405} (\bibinfo {year} {2004})}\BibitemShut {NoStop}%
\bibitem [{\citenamefont {{Froese Fischer}}\ \emph {et~al.}(2019)\citenamefont {{Froese Fischer}}, \citenamefont {Gaigalas}, \citenamefont {Jönsson},\ and\ \citenamefont {Bieroń}}]{Fischer2019GRASP}%
  \BibitemOpen
  \bibfield  {author} {\bibinfo {author} {\bibfnamefont {C.}~\bibnamefont {{Froese Fischer}}}, \bibinfo {author} {\bibfnamefont {G.}~\bibnamefont {Gaigalas}}, \bibinfo {author} {\bibfnamefont {P.}~\bibnamefont {Jönsson}},\ and\ \bibinfo {author} {\bibfnamefont {J.}~\bibnamefont {Bieroń}},\ }\href {https://doi.org/https://doi.org/10.1016/j.cpc.2018.10.032} {\bibfield  {journal} {\bibinfo  {journal} {Computer Physics Communications}\ }\textbf {\bibinfo {volume} {237}},\ \bibinfo {pages} {184} (\bibinfo {year} {2019})}\BibitemShut {NoStop}%
\end{thebibliography}%

\end{document}


\title{Supplemental Material: \\
Nuclear Deformation Effects in the Spectra of Highly Charged Ions}

\author{Zewen Sun, Igor A. Valuev, Natalia S. Oreshkina}
\affiliation{Max-Planck Institute for Nuclear Physics, Saupfercheckweg 1, 69117 Heidelberg, Germany}

\maketitle

\section{nuclear deformation corrections}
\label{sec:ND corrections} 

\begin{figure*}[h]
    \centering
    \includegraphics[width=0.95\textwidth]{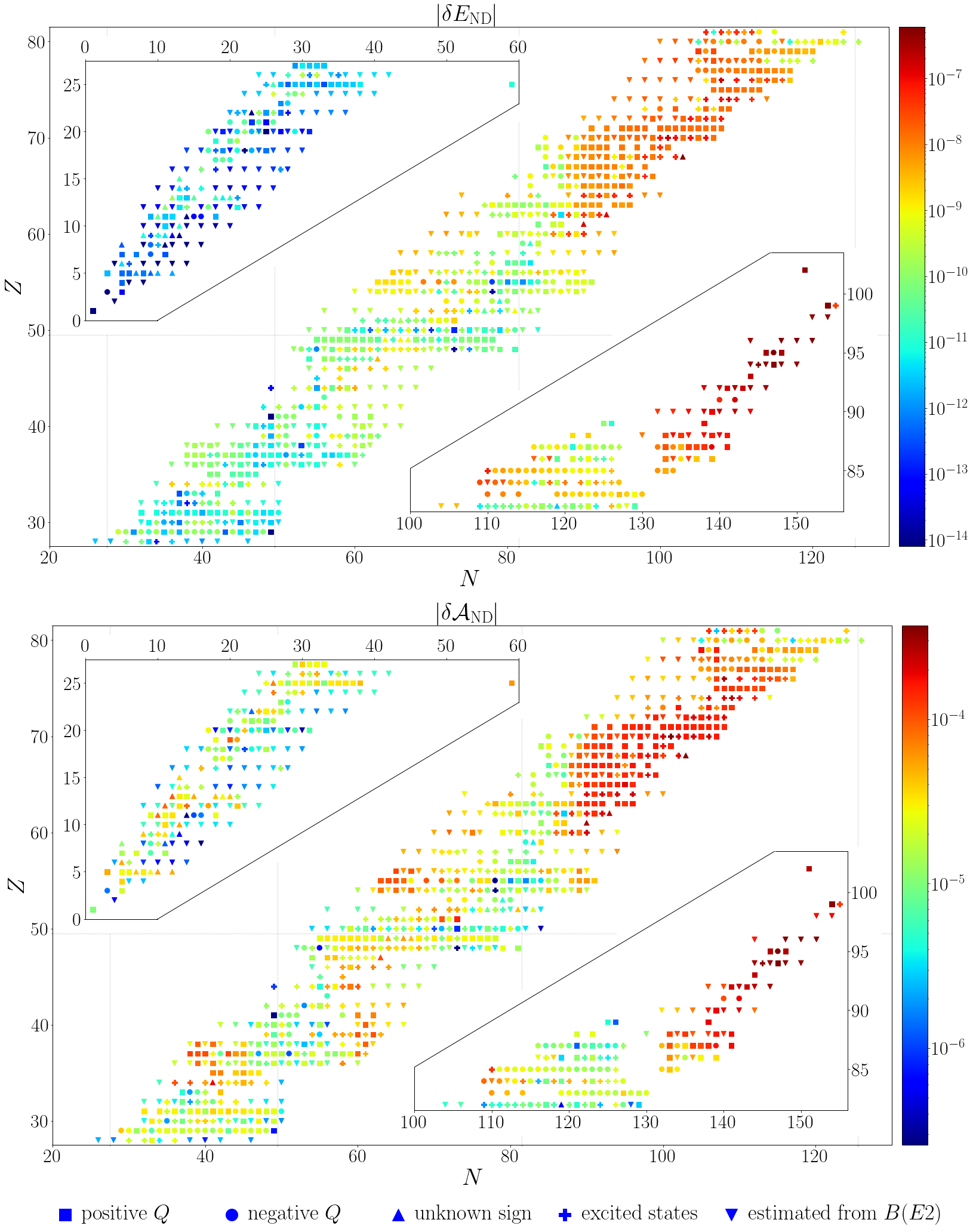} 
    \caption
    {\label{fig:S1}Nuclear chart for ND correction $\delta E_{\mathrm{ND}}$ and $\delta \mathcal{A}_{\mathrm{ND}}$ in $1s$ state. 
    The gray grid lines refer to the magic numbers on the periodic table (20, 28, 50, 82, and 126). The marker styles indicate the properties of $Q$ values used to calculate the ND correction, i.e., square: $Q$ is positive; circle: $Q$ is negative; up triangle: the sign of $Q$ is unknown but we take it as positive; cross: $Q$ obtained from a nuclear excited state instead of nuclear ground state; down triangle: $Q_0$ value is estimated from $B(E2)\uparrow$ transition rate (more explanation in Sec.~\ref{sec:results}). }
\end{figure*}

In this material, we provide the nuclear deformation (ND) corrections to eigenenergy $\delta E_{\mathrm{ND}}$ and hyperfine splitting constant $\delta \mathcal{A}_{\mathrm{ND}}$, for the ground electronic $1s$ state. 
The correction $\delta E_{\mathrm{ND}}$ is the relative difference between energies of deformed and non-deformed Fermi nuclear models
\begin{equation}
    \delta E_{\mathrm{ND}} = \frac{ E_{1s}^{(c\beta_2)} - E_{1s}^{(c0)} }{ E_{1s}^{(c0)} }, 
\end{equation}
where $E_{1s}^{(c0)}$ is the ground state eigenenergy of the non-deformed ($\beta_2 = 0$) Fermi nuclear model.

By numerically solving the Dirac equation we obtain the eigen-wavefunctions $\Psi_{n \kappa m}$, with
\begin{equation}
    \Psi_{n \kappa m} (\mathbf{r}) = \frac{1}{r} 
    \begin{pmatrix}
    G_{n\kappa} (r) \Omega_{\kappa m} (\theta, \varphi)\\
    iF_{n\kappa}(r) \Omega_{-\kappa m} (\theta, \varphi)
    \end{pmatrix}, 
\end{equation}
where $n$ is the principal quantum number, $\kappa$ is the relativistic angular momentum number, $m$ is the total magnetic number, while $G_{n\kappa}$ and $F_{n\kappa}$ are radial wavefunctions. 
For example, $1s$ state corresponds to the case of $n=1, \kappa=-1$. 
For each isotope, the Dirac equation is solved twice with the potentials defined by Fermi and deformed Fermi nuclear charge distribution, and the ND corrections are calculated based on the two results.

The ND correction to hyperfine splitting constant $\delta \mathcal{A}_{\mathrm{ND}}$ is 
\begin{equation}
\label{delta_A}
    \delta \mathcal{A}_{\mathrm{ND}} = \frac{ \mathcal{A}_{1s}^{(c\beta_2)} - \mathcal{A}_{1s}^{(c0)} }{ \mathcal{A}_{1s}^{(c0)} },
\end{equation}
where $\mathcal{A}$ is defined as
\begin{equation}
\label{HFS}
    \mathcal{A} \propto \int_0^\infty G_{n\kappa} (r) F_{n\kappa} (r) \frac{1}{r^2} \, dr. 
\end{equation}
Here, we omit the angular prefactor, since this equation of proportionality is sufficient for the analysis.

\section{numerical details}
\label{sec:results}

As for the data source, the root-mean-square (RMS) nuclear charge radii $R_{\mathrm{RMS}}$ are obtained from Ref.~\cite{ANGELI201369}. 
The nuclear quadrupole moment $Q$ data are taken from Refs.~\cite{STONE2016, STONE2021} and Ref.~\cite{PRITYCHENKO2016}. 
It should be noted that Ref.~\cite{STONE2021} (2021) is an update to Ref.~\cite{STONE2016} (2016), and we adopt the data from the latest one. 
In Ref.~\cite{STONE2021}, most of the $Q$ values have experimentally measured signs indicating the shape of nuclear charge distribution, and for those which do not, we assume them to be positive. 
In addition, some nuclei in Ref.~\cite{STONE2021} do not have available $Q$ data for nuclear ground states. 
The corresponding ND corrections are then calculated using the $Q$ from the lowest available nuclear excited state. 
The $Q_0$ values obtained from Ref.~\cite{PRITYCHENKO2016} are calculated from experimentally measured $B(E2)\uparrow$ transition rates, which implicitly assume positive $Q_0$. 
The data calculated from $Q_0$ with different signs and different references are separated on the plot by different marker styles (see Fig.~\ref{fig:S1}).

Additionally, for the spinless nuclei, the hyperfine splitting constants turn into zero, such that the ND corrections to it are not tangible.
However, we keep the corresponding $\delta \mathcal{A}_{\mathrm{ND}}$ data on Fig.~\ref{fig:S1}, defined by Eq.~\ref{delta_A} for electronic part exclusively, for the sake of completeness and since the excited nuclear states can have non-zero spin. 
Although the ND effects appear to be significant for $\delta \mathcal{A}_{\mathrm{ND}}$, for a relative change of $\sim 10^{-4}$ in heavy ions, it is difficult to factorize this ND correction from the uncertainties of finite-nuclear-size and Bohr-Weisskopf effects~\cite{SHABAEV2006109}.

Numerical accuracy of the deformation parameters are limited by the order of $10^{-16}$, and the errors can propagate to $\sim 10^{-15}$ during the calculations. 
Therefore, we do not discriminate between values below $10^{-14}$, such that the results cannot be contaminated by numerical errors.